\documentclass[preprint,showpacs,preprintnumbers,amsmath,amssymb]{revtex4}

\usepackage{graphicx}% Include figure files
\usepackage{dcolumn}% Align table columns on decimal point
\usepackage{bm}% bold math
\usepackage{subfigure}%

\begin{document}

%\preprint{APS/123-QED}

\title{Compensation of Strong Thermal Lensing in High Optical Power Cavities\\}% Force line breaks with \\

\author{C. Zhao, J. Degallaix, L. Ju, Y. Fan, D.G. Blair}
\affiliation{%
School of Physics, the University of Western Australia, 35 Stirling
Highway, Nedlands, WA 6009\
}%

\author{B.J.J. Slagmolen, M.B. Gray, C.M. Mow Lowry, D.E. McClelland}
\affiliation{%
Centre for Gravitational Physics, The Australian National
University, Canberra, 0200, Australia\
}%

\author{D. J. Hosken, D. Mudge, A. Brooks, J. Munch, P.J. Veitch}
\affiliation{%
Department of Physics, The University of Adelaide, Adelaide, South
Australia, 5005 Australia\
}%

\author{M. A. Barton, G. Billingsley }
\affiliation{%
LIGO, California Institute of Technology, Pasadena CA 91125, USA\\
}%

\date{\today}% It is always \today, today,

\begin{abstract}
In an experiment to simulate the conditions in high optical power
advanced gravitational wave detectors such as Advanced LIGO, we show
that strong thermal lenses form in accordance with predictions and
that they can be compensated using an intra-cavity compensation
plate heated on its cylindrical surface. We show that high finesse
~1400 can be achieved in cavities with internal compensation plates,
and that the cavity mode structure can be maintained by thermal
compensation. It is also shown that the measurements allow a direct
measurement of substrate optical absorption in the test mass and the
compensation plate.
\end{abstract}

\maketitle

\begin{figure}
\includegraphics{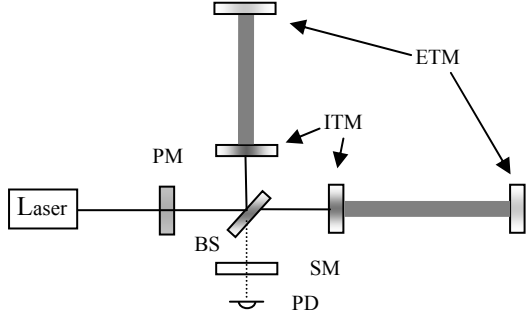}
\caption {Simplified schematic diagram of the optical layout of an
advanced interferometer; ETM: end test mass, ITM: inboard test mass,
PM: power recycling mirror, SM: signal recycling mirror, BS:
beamsplitter, PD: photodiode. Shading illustrates the thermal
lensing in the various components.}\label{fig1}

\end{figure}

Laser interferometric gravitational wave (GW) detectors employ
advanced Michelson interferometers to detect the differential strain
produced in the two arms by a passing GW. A schematic of an advanced
GW detector is shown in Fig.~\ref{fig1}\cite{b1}. Fabry-Perot
cavities are placed in the arms of the interferometer to increase
the stored power, thereby increasing the sensitivity of the
detector. The mirrors for these cavities (the inboard test mass
(ITM) mirror and end test mass (ETM) mirror) will be separated by 4
km in Advanced LIGO\cite{b1} and will have radii of curvature of
about 2 km, producing TEM00 beam radii of about 6 cm at the mirrors.
Operating the interferometer on a dark fringe and using the
signal-recycling mirror (SM) to enhance resonantly the signal
sidebands further increases the sensitivity. The power travelling
back towards the laser is resonantly reflected back into the
interferometer by the power-recycling mirror (PM), thereby reducing
the required laser power.

In advanced interferometers, the stored power will be increased
significantly to improve the detector sensitivity. In Advanced LIGO,
for example, it is expected that about 1 kW will be incident on the
back surface of the ITMs and about 830 kW may be stored in the arm
cavities. This will increase the sensitive range for neutron star
inspiral events from the present ~10 Mpc to about 200 Mpc, thus
leading to a signal event rate of many per year \cite{b2}. However,
heating due to absorption in the beam splitter (BS) and ITM
substrates, and in the reflective coatings of the ITM and ETM is
expected to cause significant wavefront distortion of the optical
modes of the interferometer \cite{b3,b4,b5,b6,b7,b8,b9}. This
distortion, also referred to as thermal lensing, would reduce the
sensitivity of the detector and could lead to instrument failure
\cite{b4}. Indeed, compensation of thermal lensing due to excess
absorption in an ITM has already found to be required in the initial
LIGO interferometer \cite{b10,b11}. In that case, a CO2 laser beam
is used to heat the back surface of the ITM to cancel the positive
lens produced by absorption \cite{b12}

The High Optical Power Test Facility (HOPTF), located at Gingin in
Western Australia, has been developed to investigate the effects of
optical absorption and its compensation in optical cavities with
high stored power \cite{b13}. It has a maximum cavity length of 80 m
and currently has a single-frequency 10 W Nd:YAG laser \cite{b14}.
Though not in use in the experiments reported here, auto-alignment
of the laser beam into the optical cavity \cite{b15} has been
achieved. An off-axis Hartmann wavefront sensor that will allow
spatially resolved, sensitive measurements of the wavefront
distortion in the ITM and CP \cite{b16} will shortly be installed.

In this letter we report on the observation and characterization of
strong thermal lensing in the sapphire ITM, caused by a distortion
that is similar in magnitude to that expected in Advanced LIGO, and
its compensation using direct heating the cylindrical surface of a
fused silica compensation plate inside the cavity. The distortion
does not cause our optical cavity to become unstable as the change
in effective curvature of the ITM is small compared to the curvature
of the ETM. In an advanced interferometer, however, this curvature
would be similar to that of the cavity mirrors.

\begin{figure}
\includegraphics{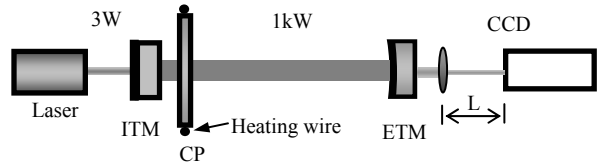}
\caption {A schematic diagram of the experiment setup. The beam from
the 10 W laser is mode matched into the Fabry-Perot cavity formed by
the back-surface ITM and the ETM. The compensation plate (CP) is
placed near the ITM and heated by 6 turns of nichrome wire insulated
by Teflon tape from the plate. The Fourier transform of the beam
transmitted through the ETM is recorded by a CCD camera placed near
the back focal plane of the 50 mm diameter, 500 mm focal length
lens}\label{fig2}

\end{figure}

Initially, we are investigating the effects of absorption in the
substrate of a sapphire ITM using a simple Fabry-Perot cavity that
has a rear-surface input-coupler. The layout of the experiment is
shown in Fig.~\ref{fig2} and the parameters of the cavity are listed
in Table~\ref{table1}. An intra-cavity Compensation Plate (CP) is
also included in the cavity. The CP is conductively heated on its
cylindrical surface and is located near the ITM. While the CP in our
experiment is directly mounted on a breadboard, the CP in an
advanced detector would be suspended and radiatively heated to avoid
noise coupling. The measured cavity lifetime was about 118 ms,
giving a finesse of about 1400. The cold-cavity beam waist was 8.7
mm.

\begin{table}
\caption{\label{table1}Parameters of the cavity}
\begin{ruledtabular}
\begin{tabular}{cccccccc}
&ITM&ETM&CP\\
\hline
Radius of Curvature (m) & Flat & 720 & Flat\\
Materials  & Sapphire  &  Sapphire & Fused Silica \\
Diameter (mm) &100 & 150 & 160\\
Thickness (mm) &46  & 80 & 17 \\
HR transmission (ppm) & 1840 100 & 20 & \\
AR reflectivity (ppm) &    29 20   & 12 12 & 100\\
Cavity internal power (kW)& 1.0\\
Cavity length (m)&77\\
\end{tabular}
\end{ruledtabular}
\end{table}

\begin{figure}
\includegraphics[width=4 in]{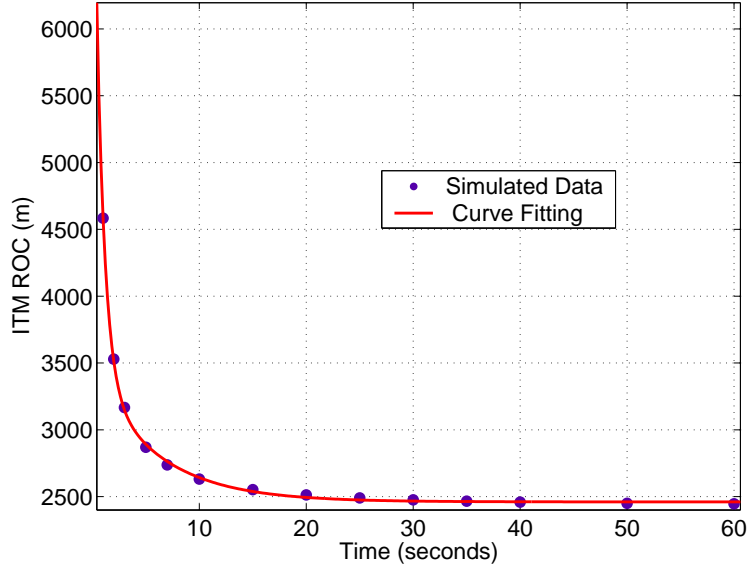}
\caption {Simulation results of ITM thermal lensing time dependence
when 0.5 W was absorbed; the dots are simulated results and the
solid line is the fitted the curve using the sum of two exponentials
with time constants of 0.76 s and 5.91 s.}\label{fig3}

\end{figure}

The expected wavefront distortion was calculated using a Finite
Element Model (FEM) and assuming an absorbed power of 0.5 W,
corresponding to a substrate absorption coefficient of 50 ppm/cm and
1 kW circulating power. The radius of curvature due to heating of
the ITM substrate is about 2.5 km, comparable to that expected for
an advanced GW interferometer using fused silica as the test mass
material \cite{b17}. The predicted time dependence of the change in
effective curvature of the ITM due to substrate absorption is shown
in Fig.~\ref{fig3}.

The time dependence can be described by the sum of two exponentials
with time constants $\tau_{1}$ and $\tau_{2}$.  The first is due to
the time constant for heating the Gaussian beam profile within the
test mass. The second characterizes the much longer time it takes
the test mass to come into equilibrium with the heated beam volume.
This is consistent with the analytical model of Hello and Vinet
\cite{b17}, which predicts an infinite set of exponentials with
rapidly decreasing time constants, of which only the longest will be
discernible.  For our sapphire ITM, we find that $\tau_{1}$ = 0.76 s
and $\tau_{2}$ = 5.91 s. It is interesting to note that the value of
the time constants  1 and  2 slightly depend on the mirror geometry
and so are only valid for the HOPTF ITM. Fig.~\ref{fig3} also shows
a comparison of the predictions of the FEM model and the exponential
fit. The thermal lensing of the fused silica compensation plate (CP)
can be similarly modelled, yielding time constants of $\tau_{1}$ =
7.40 s and $\tau_{2}$ = 50.1 s.

\begin{figure}
\includegraphics[width=4 in]{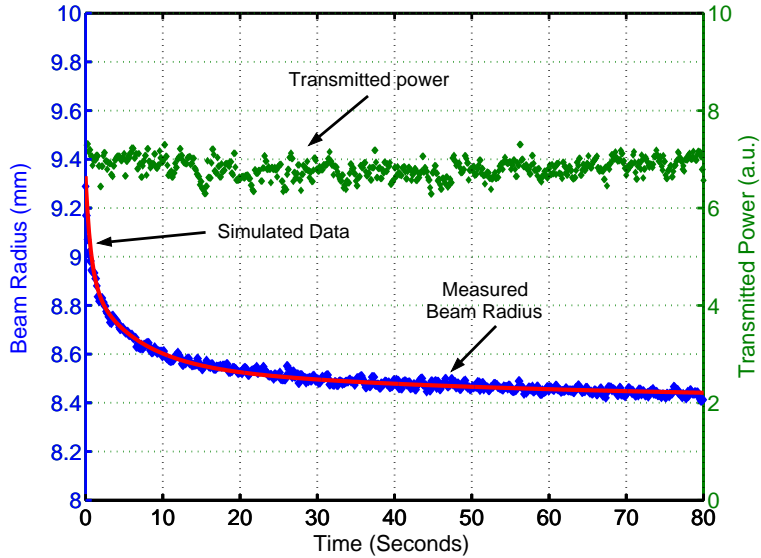}
\caption {A plot of the measured time dependence of the beam radius
at the ETM (dots) and the result of a simulation in which only the
power absorbed by the ETM and CP were optimised (solid line). The
cavity circulating power is shown on the top of the
graph.}\label{fig4}

\end{figure}

The effect of the thermal lensing and compensation was measured
using the beam transmitted through the ETM, as described in
Fig.~\ref{fig2}. The time dependence of the beam radius at the ETM,
calculated using the beam radius at the CCD, and the transmitted
power are plotted in Fig.~\ref{fig4}. The recorded spots were
accurately described by a fundamental transverse mode Gaussian
intensity profile and no obvious asymmetric distortion was observed.
The cold-cavity beam radius at the CCD was 1.0 mm, corresponding to
the beam radius of 9.3 mm at the ETM. As expected, thermal lensing
at the ITM decreases the beam size at the ETM.

Since the time constants for the thermal lensing are known, only 2
free parameters, the powers absorbed by the ITM and CP, can be used
to fit the expected time dependence to the measurements. The solid
line in Fig.~\ref{fig4} shows this time dependence for absorbed
powers of 0.51 W and 5.5 mW, respectively. The absorbed power in the
ITM corresponds to absorption of ~51 ppm/cm in the ITM.

\begin{figure}
\includegraphics[width=4 in]{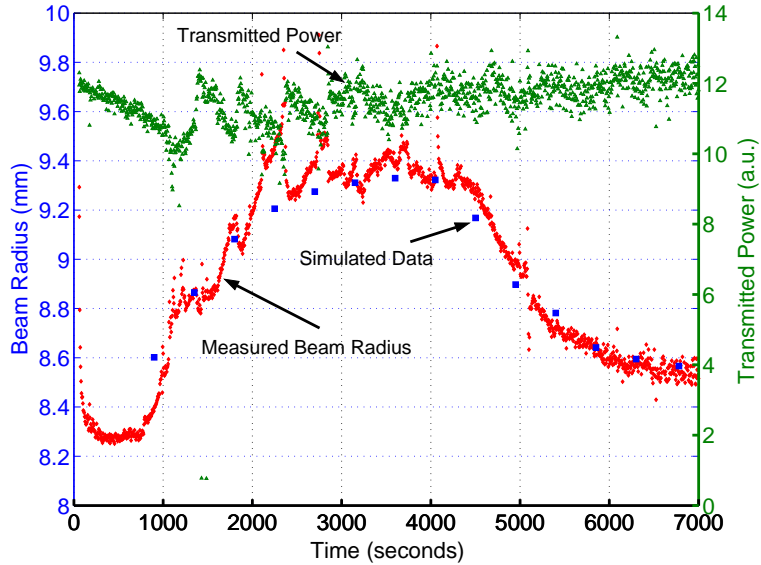}
\caption {Time dependence of the measured (dots) and simulated
(squares) ETM beam radius during thermal compensation. Optical power
storage commenced at t = 0; CP heating power was turned on at 450 s
and turned off at time 4026 s. The power transmitted through the
ETM, which is proportional to the stored cavity power, is also
shown.}\label{fig5}

\end{figure}

Compensation of the thermal lens is shown in Fig.~\ref{fig5}. The
rapid evolution of the thermal lens is just discernable at time
zero, as the ETM beam radius reduces from 9.4 mm to 8.3 mm. Heating
(9.5 W) was applied to the CP at time t = 450 s and stopped at t =
4026 s. The glitches in the measurements are due to fluctuations in
the stored power, also shown in Fig.~\ref{fig5}, caused by changes
in the alignment of the cavity and input beam.

The compensation was simulated using a model similar to that used
for the simulation of the thermal lensing with additional conduction
heating of the CP. The results, shown as squares in Fig.~\ref{fig5},
agree reasonably well with the measurements.

In conclusion, we have observed significant wavefront distortion due
to optical absorption in the substrate of a mirror, using conditions
that are traceable to those expected in Advanced LIGO. The observed
distortion is consistent with that expected from modelling, and
could lead to significant reduction in sensitivity and perhaps
instrument failure. The time evolution of thermal lensing shows a
double exponential dependence in agreement with the predictions of
our finite element model \cite{b6} and the analytical model of Vinet
and Hello \cite{b7}. Measurement of the thermal time constants allow
an accurate estimate of the test mass optical absorption. We have
also shown that the distortion can be compensated using a
conductively heated, fused silica compensation plate. Furthermore,
it appears feasible that the compensation could be maintained using
a system in which the transmitted beam size is used to define an
error signal for feedback control.

We would like to thank the International Advisory Committee of the
ACIGA/LIGO High Power Test Facility for their encouragement and
advice. This research was supported by the Australian Research
Council and the Department of Education, Science and Training and by
the U.S. National Science Foundation. It is a project of the
Australian Consortium for Interferometric Gravitational Astronomy in
collaboration with LIGO. We thank especially Barry Barish, Stan
Whitcomb and David Reitze whose support made this project possible.

\end{document}